\documentclass[preprint]{elsarticle}
\pdfoutput=1
\usepackage[T1]{fontenc}
\usepackage{subfig}
\usepackage{caption}
\usepackage{lineno,hyperref}
\usepackage{amsmath}
\usepackage{amssymb}
\usepackage{mathtools}
\usepackage{tablefootnote}
\usepackage{soul}
\usepackage{ulem}
\usepackage{color}
\usepackage{float}
\usepackage{makecell}
\usepackage[usenames,dvipsnames,svgnames,table]{xcolor}

\begin{document}
\begin{frontmatter}
\title{Can Dark Matter be Geometry? A Case Study with Mimetic Dark Matter} 

\author[ICC,UB]{Ali Rida Khalifeh\corref{cor1}}
\ead{ark93@icc.ub.edu}
\cortext[cor1]{Corresponding author}

\author[ICC,UB]{Nicola Bellomo,}
\ead{nicola.bellomo@icc.ub.edu}

\author[Baltimore,ICC,UB]{Jos\'e Luis Bernal,}
\ead{jbernal2@jhu.edu}

\author[ICC,ICREA]{Raul Jimenez}
\ead{raul.jimenez@icc.ub.edu}

\address{ICC, University of Barcelona, Marti  i Franques, 1, E-08028 Barcelona, Spain.}
\address[ICC]{ICC, University of Barcelona, Marti  i Franques, 1, E-08028 Barcelona, Spain.}
\address[UB]{Dept. de  Fisica Quantica i Astrofisica, University of Barcelona, Marti  i Franques 1, E-08028 Barcelona, Spain.}
\address[ICREA]{ICREA, Pg. Lluis Companys 23, Barcelona, E-08010, Spain.}
\address[Baltimore]{Department of Physics and Astronomy, Johns Hopkins University, 3400 North Charles Street, Baltimore, Maryland 21218, USA}

\begin{abstract}
We investigate the possibility of dark matter being a pure geometrical effect, rather than a particle or a compact object, by exploring a specific modified gravity model: mimetic dark matter. We present an alternative formulation of the theory, closer to the standard cosmological perturbation theory framework. We make manifest the presence of arbitrary parameters and extra functions, both at background level and at first order in perturbation theory. We present the full set of independent equations of motion for this model, and we discuss the amount of tuning needed to match predictions of the theory to actual data. By using the matter power spectrum and cosmic microwave background angular power spectra as benchmark observables, we explicitly show that since there is no natural mechanism to generate adiabatic initial conditions in this specific model, extra fine-tuning is required. We modify the publicly available Boltzmann code \texttt{CLASS} to make accurate predictions for the observables in mimetic dark matter. Our modified version of \texttt{CLASS} is available on GitHub\footnote{\url{https://github.com/ark93-cosmo/CLASS_Modified_MDM}}. We have used mimetic dark matter as an illustration of how much one is allowed to change the initial conditions before contradicting observations when modifying the laws of gravity as described by General Relativity. Moreover, we point out that modifying gravity without providing a natural mechanism to generate adiabatic initial conditions will always lead to highly fine-tuned models.
\end{abstract}
\end{frontmatter}

\section{Introduction}
The General Theory of Relativity (GR) has proven to be a very successful theory to describe and predict almost all of the gravitational phenomena observed to this date~\cite{will:grstatusreview}. It has, so far, been tested over a wide range of scales, ranging from the weak field regime of our solar system, all the way to cosmological scales or to the strong regime (through recent detections of black holes coalescence events~\cite{Abbott:2016blz,abbott:multimessenger,abbott:o12catalog} and most recently the super-massive black hole imaging by the Event Horizon Telescope~\cite{eht:smbhdetection}).

On the other hand, explaining the universe using GR at galactic and cosmic scales requires additional non-baryonic components. These are cold dark matter (DM) and dark energy, and the model associated with these two in GR is henceforth known as $\Lambda$CDM, where $\Lambda$ stands for the cosmological constant describing dark energy. The $\Lambda$CDM, according to state-of-the-art observational results~\cite{Aghanim:2018eyx, Alam_bossdr12}, is able to describe with astonishing precision our Universe.

This model is still a phenomenological one, and suffers from a number of conceptual problems that prevents us yet from calling it the ultimate model describing the Universe. The nature of DM itself is still unknown, despite of the many DM candidates that have been proposed, including particles, compact objects and gravity effects, see e.g., Refs.~\cite{Bertone:2016nfn, bertone:dmstatusreview} and Refs. therein. Typical modifications of GR have been driven by the presence of unsolved problems, and they typically address the issue of describing the dark energy sector by introducing a scalar field~\cite{Ratra}. Since GR is extrapolated from solar system scales up to cosmological scales, it is not impossible that what we currently interpret as DM is in reality a pure gravitational effect.
 
In this work we analyse the Mimetic Dark Matter (MDM) model~\cite{chamseddine:mimeticdm}, in which DM, instead of being made of particle or compact objects, is described by gravity. In its original form, the model was a reformulation of GR, whereby the physical metric was rewritten in terms of an auxiliary one and a scalar field, while maintaining the language of differential geometry. The authors of Ref.~\cite{chamseddine:mimeticdm} claim that GR is already able to describe DM without explicitly adding pressureless dust particles to the energy content of the universe: the scalar field is not a new dynamical degree of freedom, as stated in the footnote of Ref.~\cite{chamseddine:mimeticdm}. The model has been investigated thoroughly: alternative formulations have been provided~\cite{Golovnev:lagrangeMultiplier}, the absence of ghosts has been proven~\cite{Barvinsky:ghost, Ganz:2018mqi}, it has been generalized to explain other cosmological phenomena, as inflation or dark energy~\cite{chamseddine:mimeticcosmo}, and to Horndeski theories of gravity~\cite{arroja:mimeticdisformallagrange, arroja:mimeticcosmoperturbations, arroja:mimeticlss, Ganz:2018vzg}. We refer the interested reader to Ref.~\cite{Sebastiani:2016ras} and refs. therein to a complete discussion of different aspects of MDM. 

However, some aspects of this theory remain unexplored, in particular the degree of tuning necessary to match the theory with current observations. In this work, we re-examine the derivation of the equations of motion for MDM, and show that the redundancy in the latter still provides the equivalent of Friedmann equations, but at the expense of new arbitrary functions that require tuning to match the observed Universe. Moreover, since the standard MDM model does not include a mechanism able to produce adiabatic initial conditions, further tuning is required\footnote{This issue was already  mentioned in Ref.~\cite{chamseddine:mimeticdm}, and was also discussed in Refs.~\cite{mirzagholi:imperfectdarkmatter, ramazanov:imperfectdarkmatter}. In the latter, adiabaticity was recovered by introducing extra specific functions.}. In particular, we show that future cosmic-variance limited experiments are able to detect even a slight departure from adiabaticity, constraining the model unless some degree of fine tuning is assumed.

The paper is organized as follows: in section~\ref{sec:mimetic_gravity} we review the mimetic dark matter model. In section~\ref{sec:einstein_equations} we describe the structure of the model and its equations of motion, both at the background level and at first order in perturbation theory, describing the freedom in the choice of particular parameters of the theory. In section~\ref{sec:ICandObservations} we investigate the observational consequences of this extra freedom, and we discuss the degree of tuning required by the theory. Finally we conclude in section~\ref{sec:conclusions}.

Throughout this work, we use the $(-,+,+,+)$ signature and units in which $\hbar=c=1$. Spacetime indices are denoted by Greek letters  and range from $0$ to $3$, while spatial indices are denoted by Latin letters and range from $1$ to $3$; repeated indices are summed over.


\section{The Mimetic Dark Matter Model}
\label{sec:mimetic_gravity}
The original MDM model proposed a reformulation of the physical metric $g_{\rho\sigma}$ as\footnote{Note that there will be sign differences compared to~\cite{chamseddine:mimeticdm}and~\cite{chamseddine:mimeticcosmo} because we use an opposite metric signature.}~\cite{chamseddine:mimeticdm}
\begin{equation}
g_{\rho\sigma}=-\frac{1}{\mu^4}\bigg(\tilde{g}^{\alpha\beta}\partial_{\alpha}\varphi\partial_{\beta}\varphi\bigg)\tilde{g}_{\rho\sigma},
\label{eq:FirstMDMmetric}
\end{equation}
where $\tilde{g}_{\rho\sigma}$ is an auxiliary metric, $\varphi$ is a scalar field, called mimetic field, and $\mu$ is an arbitrary factor we introduced here just to make explicit the freedom in rescaling the mimetic field by a constant factor. The physical metric corresponds to the auxiliary metric multiplied by a conformal factor that depends on the latter. In this way, one is said to express explicitly the conformal mode of gravity, which is manifested by the invariance of the physical metric under a conformal transformation of the auxiliary one. This conformal mode is now encoded in the scalar field~$\varphi$, and therefore the auxiliary metric will no longer be used (see, e.g., Ref.~\cite{Barvinsky:ghost} for more details on the conformal mode of gravity in these models). The mimetic field has to obey the so called constraint equation
\begin{equation}
g^{\rho\sigma}\partial_{\rho}\varphi\partial_{\sigma}\varphi + \mu^4 = 0,
\label{eq:FirstMDMconstraint}
\end{equation}
obtained by contracting the mimetic field derivatives~$\partial_{\rho}\varphi\partial_{\sigma}\varphi$ with the inverse of the physical metric, which reads as $g^{\rho\sigma}=P^{-1}\tilde{g}^{\rho\sigma}$, where $P=(\tilde{g}^{\alpha\beta}\partial_{\alpha}\varphi\partial_{\beta}\varphi)/\mu^4$.

The Einstein-Hilbert action reads as~\cite{chamseddine:mimeticdm}
\begin{equation}
S = \int d^{4}x \sqrt{-g(\tilde{g},\varphi)}\bigg[\frac{M_p^2}{2}R(g_{\mu\nu}(\tilde{g}_{\mu\nu},\varphi))+ \mathcal{L}_{mr}[g_{\mu\nu},\psi_m,\psi_r]\bigg],
\end{equation}
where $g$ is the determinant of the physical metric, $M_p=(8\pi G)^{-1/2}$ is the reduced Planck mass, $R$ is the Ricci scalar and $\mathcal{L}_{mr}$ is the Lagrangian that describes baryonic matter and radiation fields $\psi_m$ and $\psi_r$, respectively. By varying the action with respect to the auxiliary metric, we get Einstein equations of the form:
\begin{equation}
M_p^2G^{\mu\nu}=T^{\mu\nu}+\tilde{T}^{\mu\nu},
\label{eq:einstein_equations_original_mdm}
\end{equation}
where $G_{\mu\nu}=R_{\mu\nu}-(1/2)g_{\mu\nu}R$ is the Einstein tensor defined in terms of the Ricci tensor~$R_{\mu\nu}$ and the Ricci scalar~$R$, $T_{\mu\nu}=-(2/\sqrt{-g})\delta S_{mr}/\delta g^{\mu\nu}$ is the stress-energy tensor of the baryonic matter and radiation fields, with $S_{mr}$ being their corresponding action, and $\tilde{T}^{\mu\nu}$ takes the form of the stress-energy tensor of dust~\cite{chamseddine:mimeticdm}. Note also that a dynamical variable is any quantity by which the action is varied, which means that the auxiliary metric, as well as the physical metric, are dynamical variables. As can be seen from~\eqref{eq:einstein_equations_original_mdm}, the model predicts that the effects of DM, encoded in~$\tilde{T}^{\mu\nu}$, can be generated without the need of adding extra species to the action. Moreover, as shown in Refs.~\cite{Golovnev:lagrangeMultiplier, Barvinsky:ghost}, the MDM constraint~\eqref{eq:FirstMDMconstraint} can be incorporated in the action with the use of a Lagrange multiplier~$\lambda$. In addition, MDM can be modified (hence it becomes a modification of GR) by introducing into the action a potential for the scalar field, as has been first done in Refs.~\cite{chamseddine:mimeticcosmo, lim:constraintequation}\footnote{Notice that the mimetic field can be rescaled as $\varphi\rightarrow|\mu|\varphi$, hence it is possible to absorb the factor~$\mu$ into the Lagrange multiplier $\lambda$ at the level of the action. However, in the case of non-zero potential, the explicit form of the action may change under such rescaling.}. In that case, the action becomes\footnote{It is interesting to note that using the same action as in~\cite{chamseddine:mimeticdm,chamseddine:mimeticcosmo} with the current signature will result in an imaginary field. However the study of this case is beyond the scope of this work.}
\begin{equation}
S = \int d^4x\sqrt{-g}\left[\frac{M_p^2}{2}R(g_{\mu\nu})-V(\varphi)-\lambda\left(g^{\mu\nu}\partial_{\mu}\varphi\partial_{\nu}\varphi+\mu^4\right) + \mathcal{L}_{mr}[g_{\mu\nu},\psi_m,\psi_r]\right],
\label{eq:OurAction}
\end{equation}
where $V(\varphi)$ is a potential for the mimetic field. As done previously in~\cite{chamseddine:mimeticcosmo}, the variation of the action in equation~\eqref{eq:OurAction} with respect to the physical metric gives the new Einstein equations which, after using its trace to substitute for $\lambda$, gives
\begin{equation}
		M_p^2G_{\mu \nu }-T_{\mu\nu} = -g_{\mu \nu }V(\varphi) + \left(T - M_p^2G - 4V \right) \frac{\partial_{\mu}\varphi\partial_{\nu}\varphi - \frac{1}{2}g_{\mu\nu}\mathcal{C}}{\mathcal{C}+\mu^4},
		\label{eq:einstein_equations}
		\end{equation}
where $T=T^\mu_{\ \mu}$ and $G=G^\mu_{\ \mu}$ are the traces of $G_{\mu\nu}$ and $T_{\mu\nu}$, respectively, and 
\begin{equation}
		\mathcal{C} = g^{\alpha\beta}\partial_{\alpha}\varphi\partial_{\beta}\varphi+\mu^4.
		\label{eq:C_constraint}
		\end{equation} The constraint equation~\eqref{eq:FirstMDMconstraint} for the mimetic field, which incidentally can be derived by varying the action with respect to the Lagrange multiplier $(\delta S/\delta\lambda=0)$, is equivalent to $\mathcal{C} = 0$. Moreover, we will impose the constraint equation when we derive the Friedmann equations at the background level in the next section. This will allow us to track the quantities that are affected by it. The appearance of an opposite sign in equation~\eqref{eq:einstein_equations} compared to equation (2.4) of Ref.~\cite{chamseddine:mimeticcosmo} is due to our opposite choice of the metric signature. Moreover, if we compare the RHS of equation~\eqref{eq:einstein_equations} to the stress-energy tensor of a perfect fluid:
\begin{equation}
\tilde{T}^\mu_{\ \nu} = pg^\mu_{\ \nu} + (\rho+p)u^\mu u_\nu,
\label{eq:stress_energy_tensor}
\end{equation}
where $\rho$ is the energy density, $p$ is the pressure and $u^{\mu}=dx^{\mu}/\sqrt{-ds^2}$ is the corresponding $4$-velocity of the fluid, we see that we can re-obtain equation~\eqref{eq:einstein_equations_original_mdm} by identifying as energy density, pressure and $4$-velocity of the scalar field fluid, respectively with
		\begin{align}
		&\rho_{\varphi}=\big(1+\frac{1}{2}\mathcal{C}\big)\big(T-M_p^2G-4V(\varphi)\big)+V;\nonumber
		\\ &p_{\varphi}=-\frac{1}{2}\mathcal{C}\big(T-M_p^2G-4V(\varphi)\big)-V(\varphi);\nonumber
		\\
		& u^{\mu}=\partial^{\mu}\varphi/\sqrt{\mathcal{C}+\mu^4}.
		\label{eq:MimeticStressTensorProperties}
		\end{align}
		
In this case equation~\eqref{eq:FirstMDMconstraint} corresponds to the normalization equation $u^{\mu}u_{\mu}=-1$ for the $4$-velocity. The equations of dynamics for matter and radiation fields $(\delta S/\delta\psi_m, \delta S/\delta\psi_r=0)$ do not change with respect to the standard ones in GR. Moreover, the stress-energy tensor of matter and radiation fields is conserved as in GR, namely we have $\nabla_\mu T^{\mu\nu}=0$, even if Einstein equations~\eqref{eq:einstein_equations} changed. For this reason we do not report them here and we refer the interested reader to Ref.~\cite{ma:cosmoperturbations}.


\section{Einstein Equations}
\label{sec:einstein_equations}
In the following we specify the general equations of the MDM model to the case of an Universe described by Friedman-Lema\^{i}tre-Robertson-Walker metric. Since we are interested only in the description of scalar modes, we choose to work with the conformal Newtonian gauge described by the metric
\begin{equation}
ds^2 = g_{\mu\nu}dx^\mu dx^\nu = a^2(\tau)\left[-(1+2\Psi)d\tau^2 + (1-2\Phi)\delta_{ij}dx^idx^j\right],
\label{eq:TheMetric}
\end{equation}
where $a$ is the scale factor, $\tau$ is the conformal time, $x^j$ are comoving spatial coordinates and the potentials $\Psi$ and $\Phi$ are related to Bardeen gauge-invariant variables~\cite{bardeen:potentials}. Throughout this section we follow notation and conventions of Ref.~\cite{ma:cosmoperturbations}.

We assume that the matter and radiation content of the Universe can be described by an almost perfect fluid with stress-energy tensor given by
\begin{equation}
T^\mu_{\ \nu} = pg^\mu_{\ \nu} + (\rho+p)u^\mu u_\nu+\Sigma^{\mu}_{\ \nu},
\label{eq:stress_energy_tensor2}
\end{equation}
where $\Sigma^\mu_{\ \nu}$ contributes to the anisotropic stress only at first order in perturbation theory. Assuming that the fluid has some small density and pressure fluctuations $\delta\rho$ and $\delta p$, coordinate velocity $v^i=dx^i/d\tau$ and anisotropic stress $\Sigma^i_{\ j}$ (such that $\Sigma^i_{\ i}=0$), the components of the stress-energy tensor, up to first order in perturbation theory, can be written as
\begin{equation}
T^0_{\ 0} = -(\bar{\rho}+\delta\rho), \qquad T^i_{\ 0} = - (\bar{\rho}+\bar{p})v^i, \qquad T^i_{\ j} = (\bar{p}+\delta p)\delta^i_{\ j} + \Sigma^i_{\ j},
\end{equation}
where an over-bar denotes background quantities. In the following we use also the overdensity contrast $\delta=\delta\rho/\bar{\rho}$, the divergence of the velocity~$\theta$ and of the traceless anisotropic stress~$\sigma$, which read as
\begin{equation}
(\bar{\rho}+\bar{p})\theta = ik^j\delta T^0_{\ j} = i(\bar{\rho}+\bar{p})k^jv_j, \qquad (\bar{\rho}+\bar{p})\sigma = -\left(\hat{k}_i\hat{k_j}-\frac{1}{3}\delta_{ij}\right)\Sigma^i_{\ j}.
\end{equation}


\subsection{Background Evolution}
\label{subsec:background_equations}
At the background level, energy densities $\bar{\rho}(\tau)$, pressures $\bar{p}(\tau)$ and the scalar field $\bar{\varphi}(\tau)$ are only time-dependent. The constraint equation~\eqref{eq:FirstMDMconstraint} reads, after fixing it for the appropriate metric signature, as
\begin{equation}
\mu^4-\frac{\bar{\varphi}'^2}{a^2} = 0,
\label{eq:constraint_equation_background}
\end{equation}
where ``$\ '\ $'' denotes derivative with respect to the conformal time $\tau$. As in GR, and as done in Ref.~\cite{chamseddine:mimeticcosmo}, Friedman equations are obtained from the $(0-0)$ and the trace of the spatial $(i-j)$ components of equation~\eqref{eq:einstein_equations} and read as
\begin{equation}
		\mathcal{A}=-\frac{1}{\mathcal{C}+\mu^4}\bigg[3\mathcal{B}-\mathcal{A}\bigg]\bigg[a^{-2}\bar{\varphi}'^2+\frac{1}{2}\mathcal{C}\bigg],
		\label{eq:identity_background}
		\end{equation}
		\begin{equation}
		\mathcal{B}=\frac{\mathcal{C}}{2(\mathcal{C}+\mu^4)}\bigg[3\mathcal{B}-\mathcal{A}\bigg]
		\label{eq:spatial_einstein_equation_background},
		\end{equation}
		where
		\begin{equation}
		\mathcal{A}=3M_p^2\mathcal{H}^2-a^2\bar{\rho}-a^2V,\quad \mathcal{B}=-M_p^2\big(2\mathcal{H}'+\mathcal{H}^2\big)-a^2\bar{p}+a^2V
		\end{equation}
 and $\mathcal{H}=a'/a$ is the Hubble expansion parameter in conformal time. Note that these two equations are identical once the explicit form of $\mathcal{C}$, eq.\eqref{eq:C_constraint}, is substituted. Therefore, the final form of either of them would be:
\begin{equation}
\big(a^2\mu^4-\bar{\varphi}'^2\big)\mathcal{A}+\big(a^2\mu^4+\bar{\varphi}'^2\big)\mathcal{B}=0.
\label{eq:After_C}
\end{equation} 
If we now impose the constraint equation~\eqref{eq:constraint_equation_background} to~\eqref{eq:After_C}, in order to make it consistent, we deduce that:
\begin{equation}
\mathcal{B}=0\quad\Rightarrow\quad M_p^2\big(2\mathcal{H}'+\mathcal{H}^2\big)+a^2\bar{p}-a^2V=0
\label{eq:Second_Friedmann}
\end{equation}
and
\begin{equation}
\mathcal{A}=f(\tau)\quad\Rightarrow\quad 3M_p^2\mathcal{H}^2 - a^2\bar{\rho} - a^2V = f(\tau).
\label{eq:f_definition}
\end{equation}
These two equations have the same form as the 2${}^{\text{nd}}$ and 1${}^{\text{st}}$ Friedmann equations, respectively. Therefore, although we have a redundancy at the level of Einstein equation's components, we still get the Friedmann equations. However, this is at the expense of getting an arbitrary function of conformal time, $f(\tau)$, which connects the expansion history of the universe to its energy. Indeed, as done before, if we identify the energy density of the scalar field with $\bar{\rho}_\varphi = a^{-2}f(\tau) + V$, equations~\eqref{eq:MimeticStressTensorProperties}, \eqref{eq:Second_Friedmann} and~\eqref{eq:f_definition} give the following definitions for the background density and pressure of the field, respectively:
\begin{equation}
\bar{\rho}_{\varphi}=T - M_p^2G - 3V=f(\tau)a^{-2}+V;\quad \bar{p}_{\varphi} = -V.
\label{eq:definitions_p_rho_phi}
\end{equation}
 
Note that $f(\tau)$ function can be determined by taking the time derivative of equation~\eqref{eq:f_definition}, using the conservation equation, $\bar{\rho}' + 3\mathcal{H}(\bar{\rho}+\bar{p}) = 0$, and comparing the result to equation~\eqref{eq:Second_Friedmann}. We find that $f(\tau)$ has to satisfy the differential equation
\begin{equation}
f'+\mathcal{H}f + \left[(a^2V)'-2\mathcal{H}(a^2V)\right] = 0.
\label{eq:f_V_differential_equation}
\end{equation}
The solution of the homogeneous equation reads $f=\kappa / a$, where $\kappa$ is a space-independent integration constant because of homogeneity and isotropy. The general solution depends on the shape of the potential and is given by
\begin{equation}
f(\tau) = \frac{\kappa}{a} - a^2V(\varphi) + \frac{3}{a}\int^\tau_{\tau_0} d\tilde{\tau} \left(\mathcal{H}a^3V\right),
\label{eq:general_solution_f}
\end{equation}
where $\tau_0$ is some reference time, and which generalises the result presented in Ref.~\cite{chamseddine:mimeticcosmo} to an Universe filled with matter and radiation. Therefore, independently of the chosen shape of the potential, a fraction of the mimetic field energy density scales as $\kappa a^{-3}$, i.e., as DM would do. The integration constant,~$\kappa$, is an extra free parameter of the theory that has to be chosen properly to fit the observation. Notice that this additional parameter is not connected to any parameter in the action of the theory, hence apart from setting its value using current observational data, we cannot assign it any value motivated by the theory itself. Therefore, when it comes to the amount of DM in the universe, the model presents at least the same level of tuning needed as in $\Lambda$CDM. We discuss how this parameter was linked to the initial conditions of our Universe in section~\ref{sec:ICandObservations}.

We conclude by highlighting the solutions for two special cases: a zero potential $(V\equiv 0)$ and a strictly negative constant potential $(V\equiv\bar{V}<0)$. In the first case we have that
\begin{equation}
\bar{\rho}_\varphi = f a^{-2} + V = \kappa a^{-3}, \qquad \bar{p}_\varphi = -V \equiv 0,
\end{equation}
hence the mimetic field plays only the role of cold DM. In the second case we have both DM and dark energy at the same time, in fact $f(\tau)=\left[\kappa+\bar{V}a^3_0\right]/a = \tilde{\kappa}/a$, where~$\tilde{\kappa}$ is a new integration constant and $a_0=a(\tau_0)$, therefore
\begin{equation}
\bar{\rho}_\varphi = f a^{-2} + \bar{V} = \tilde{\kappa} a^{-3} + \bar{V}, \qquad \bar{p}_\varphi = -\bar{V},
\label{eq:quartessence_behaviour}
\end{equation}
hence in this scenario the mimetic field plays the role both of DM and a cosmological constant, as in certain quartessence models~\cite{lima:quartessence} and as also shown in Ref.~\cite{lim:constraintequation}. In both cases, see, e.g., Ref.~\cite{chamseddine:mimeticcosmo}, we can connect the energy density and pressure of the mimetic field to the background value of the Lagrange multiplier~$2\bar{\lambda} = \bar{\rho}_\varphi + \bar{p}_\varphi = fa^{-2}$. Therefore the ambiguity in choosing the free parameter~$\kappa$ derives from the possibility of rescaling arbitrarily the mimetic field while reabsorbing any constant in the Lagrange multiplier. 


\subsection{Perturbative Dynamics}
\label{subsec:perturbation_equations}
In order to compare the predictions of this model to those of the traditional $\Lambda$CDM we need to compute the evolution of perturbations. In this section we present the equations of dynamics for the metric and scalar field fluctuations, while, as previously stated, we do not report those of matter and radiation fields since they are identical to those in GR. Moreover, the same machinery used to derive~\eqref{eq:Second_Friedmann} and~\eqref{eq:f_definition} applies to their perturbed equivalents, and therefore it will not be presented explicitly, rather simply the final results will be. Also, to simplify the expressions, we will impose the constraint equation~\eqref{eq:constraint_equation_background}, or equivalently $\mathcal{C}=0$, at the level of the perturbation of Einstein equations~\eqref{eq:einstein_equations}. The result will be the same whether we take $\mathcal{C}=0$ at this level or at very end. The results of this section apply to MDM models with an arbitrary shape of the potential, hence they can be applied to different scenarios.

The constraint equation~\eqref{eq:FirstMDMconstraint} for the mimetic field fluctuation reads as
\begin{equation}
\delta\varphi' - \bar{\varphi}'\Psi = 0,
\label{eq:var_phi}
\end{equation}
however, as done in Ref.~\cite{arroja:mimeticlss}, it is more convenient to introduce a new variable $v_\varphi=-\delta\varphi/\bar{\varphi}'$ whose equation of motion, invariant under rescaling of the mimetic field, is 
\begin{equation}
v'_\varphi + \mathcal{H}v_\varphi + \Psi = 0.
\label{eq:phi_velocity_equation}
\end{equation}
Moreover, by defining the velocity divergence of the scalar field fluid as $\theta_\varphi = k^2v_\varphi$, we find that equation~\eqref{eq:phi_velocity_equation} can be recast as
\begin{equation}
\theta'_\varphi + \mathcal{H}\theta_\varphi + k^2\Psi = 0,
\end{equation}
which is the equation of motion of the velocity divergence for a non-relativistic and collisionless fluid. Therefore the theory itself, independently of the shape of the potential, is able to reproduce the equation of motion of DM velocity divergence.

At first order in perturbation theory we have four independent Einstein equations. In reporting these equations, we keep on the LHS of each equation all the terms unchanged with respect to the GR case, see e.g., Ref.~\cite{ma:cosmoperturbations}, while on the RHS we put the new terms given by the MDM model.

By defining the pressure perturbation of the scalar field fluid as $\delta p_{\varphi}=v_{\varphi}\bar{\varphi}'V,_{\varphi}$, where $V,_{\varphi}=\partial V/\partial\varphi$, we find that the traceless part and the trace of the $(i-j)$ components of Einstein equations are given by
\begin{equation}
k^2\left(\Phi-\Psi\right) - \frac{3a^2}{2M_p^2}\left(\bar{\rho}+\bar{p}\right)\sigma = 0,
\label{eq:spatial_traceless_perturbed_einstein_equation}
\end{equation}
\begin{equation}
\Phi'' + \mathcal{H}\left(\Psi'+2\Phi'\right) + \left(2\mathcal{H}'+\mathcal{H}^2\right)\Psi + \frac{k^2}{3}\left(\Phi-\Psi\right) - \frac{a^2}{2M_p^2}\delta p = \frac{a^2}{2M_p^2}\delta p_{\varphi}.
\label{eq:spatial_trace_perturbed_einstein_equation}
\end{equation}
Notice that the mimetic field cannot be a source of anisotropic stress, i.e., $\sigma_{\varphi}\equiv 0$, and only when the gradient of the potential ($\partial V/\partial\varphi$) is non-zero the scalar field develops an isotropic pressure perturbation~$\delta p_{\varphi}$.

Using Friedman equations~\eqref{eq:Second_Friedmann} and~\eqref{eq:f_definition}, we find that the $(0-i)$ components of Einstein equations reads
\begin{equation}
\Phi'+\mathcal{H}\Psi - \frac{a^2}{2M_p^2k^2}\left(\bar{\rho}+\bar{p}\right)\theta = \frac{a^2}{2M_{p}^2k^2} (\bar{\rho}_{\varphi}+\bar{p}_{\varphi})\theta_{\varphi},
\label{eq:spatial_time_perturbed_einstein_equation}
\end{equation}
which shows that not only in MDM we have an equation for velocity divergence identical to that of DM, but also this velocity contribution appears in the correct form in Einstein equations. 
 
Finally, using the results we have found both in \S~\ref{subsec:background_equations} and in \S~\ref{subsec:perturbation_equations} so far, we are able to write the perturbed equivalent of the 1${}^{\text{st}}$ Friedmann equation as:
\begin{equation}
3\mathcal{H}\Phi' + 3\mathcal{H}^2\Psi + k^2\Phi + \frac{a^2}{2M_p^2}\bar{\rho}\delta =- \frac{a^2\bar{\rho}_{\varphi}}{2M_p^2}g(k,\tau),
\label{eq:g_definition}
\end{equation}
where this specific form has been chosen for later convenience. It is important to emphasize again the fact that even if we have redundancy in the equations of motion, we can still find the equivalence of the perturbed Friedmann equations for this model, at the expense of getting an arbitrary function, which at 1${}^{\text{st}}$ order in perturbation is $g(k,\tau)$. We argue that there is one of these functions at every level in perturbation theory, hence going to second order we would find an $h(k,\tau)$ arbitrary function, and so on.

By taking time derivatives of~\eqref{eq:g_definition} and using the results of \S~\ref{subsec:background_equations} and \S~\ref{subsec:perturbation_equations}, we get
\begin{equation}
g'=-(1+\omega_{\varphi})(\theta_{\varphi}-3\Phi')-3\mathcal{H}(c_{({\varphi}),s}^2-\omega_{\varphi})g
\label{eq:diff_eq_g}
\end{equation}
where $\omega_{\varphi}=\bar{p}_{\varphi}/\bar{\rho}_{\varphi}$ is the equation of state of the scalar field fluid, and $c_{(\varphi),s}^2=\delta p_{\varphi}/\delta\rho_{\varphi}$ is an effective sound-speed-like term of the mimetic field. If we compare equation~\eqref{eq:diff_eq_g} to equation $(30)$ of Ref.~\cite{ma:cosmoperturbations}, we notice that the function~$g$ evolves as the overdensity contrast of a fluid, therefore it can be thought as~$g\equiv\delta_{\varphi}$. In the case of zero potential, i.e., in the case where the mimetic field describes DM, we recover the evolution equation for dust.


\section{Initial Conditions and Observational Constraints}
\label{sec:ICandObservations}
We have shown that the MDM model, for any given potential~$V$, shows a level of flexibility in the choice of the function~$f(\tau)$ and in the choice of the initial conditions (ICs) for the mimetic field fluctuation and the function~$g$ at first order in perturbation theory. Notice that generalisations of the standard MDM model seems to be able to produce adiabatic ICs~\cite{mirzagholi:imperfectdarkmatter, ramazanov:imperfectdarkmatter}, however this feature depends on the form of the mimetic Lagrangian and it is not a general property of the MDM model.

At the background level, independently of the shape of the potential, we have a free parameter,~$\kappa$, whose value in principle is set by ICs of the mimetic field and cannot be derived directly from the action written in equation~\eqref{eq:OurAction}. The authors of Ref.~\cite{chamseddine:mimeticdm} suggested that if the mimetic field is coupled to the inflaton, a non-vanishing amount of DM can survive~$60$ e-folds of expansion without spoiling inflationary dynamics. Since the evolution of the mimetic field is fixed by the constraint equation~\eqref{eq:FirstMDMconstraint}, which does not provide any attractor solution, the dynamics of the inflaton need to be fine-tuned to provide the correct amount of DM today. Moreover, in case the potential is non-zero, its shape and parameters have to be tuned to match observations, as we can see in the example in equation~\eqref{eq:quartessence_behaviour}.

At the perturbation level, even if the $g$ function and the mimetic field fluctuation evolve as an energy overdensity and a velocity, respectively, we have to set ICs for both quantities. This requires a second tuning because we know them to be adiabatic~\cite{peiris:wmapng}. If the mimetic field is just a spectator field during inflation, as suggested above to fix the value of~$\kappa$, then its presence can result in having also isocurvature ICs~\cite{Padilla:2019fju}, which are largely ruled out~\cite{akrami:planckng2018}. However, having isocurvature initial conditions is not inevitable for MDM, rather it is a possibility without some level of fine-tuning. Moreover, even assuming no isocurvature ICs are generated, we still have to properly define the scalar field ICs, or, equivalently, to define an inflationary scenario able to generate adiabatic ICs also for the mimetic field.

To study the impact of ICs in this model, we consider the case study of MDM accounting only for DM, i.e., we consider the $V\equiv 0$ case. We fix~$\kappa$ to be the observed DM energy density today, we fix the ICs for~$g$ so that adiabaticity is preserved. We let vary only the IC of the mimetic field perturbation, i.e., the IC of the mimetic field fluid velocity divergence~$\theta_\varphi$, which in the adiabatic case is related to the gravitational potential by $\theta_\varphi = \frac{1}{2}(k^2\tau)\Psi$~\cite{ma:cosmoperturbations}. Variations in~$\kappa$ and ICs for~$g$, would result in larger departures from the $\Lambda$CDM case, hence we can consider our approach as conservative.
 
We modify the public code \texttt{CLASS}~\cite{blas:class} to include the effects of the MDM model accounting for DM. We parametrise deviations from standard adiabatic ICs for the velocity divergence as
\begin{equation}
\theta_\varphi = \Big[1+\alpha\sin(\log_2k)\Big]\frac{1}{2}(k^2\tau)\Psi,
\label{eq:ThetaICdeviation}
\end{equation}
where~$\alpha$ represents the maximum amplitude of the deviation from adiabatic initial conditions. Our choice in equation ~\eqref{eq:ThetaICdeviation} has been made only for illustrative purposes, to make the plots clearer and easier to be interpreted; any other small deviation from adiabatic initial conditions would be equally valid to prove our point. This choice allows us to have variations between~$\left[1-\alpha,1+\alpha\right]$ with respect to adiabatic ICs. We check that for other wavenumber dependences, for instance randomly choosing a number in~$\left[1-\alpha,1+\alpha\right]$, our findings are unchanged. In the following we assume the Planck18 baseline cosmology assuming the best-fit parameters to the whole Planck dataset~\cite{Aghanim:2018eyx}: $\omega_\mathrm{b}=0.0224$ is the physical baryon density today, $\omega_\mathrm{cdm}=0.120$ is the physical cold dark matter density today, $h=0.674$ is the reduced Hubble expansion rate today, $10^{9}A_s=2.101$ is the amplitude of the primordial scalar perturbations, $n_s=0.965$ is the scalar spectral index and $\tau=0.054$ is the optical depth to reionization. We use~$\omega_\varphi=\kappa h^2=0.120$, where $\omega_\varphi$ is the physical density of the mimetic field today, instead of $\omega_\mathrm{cdm}$ when computing observables in MDM.

\begin{figure}[t]
\centerline{
\includegraphics[width=1.0\columnwidth]{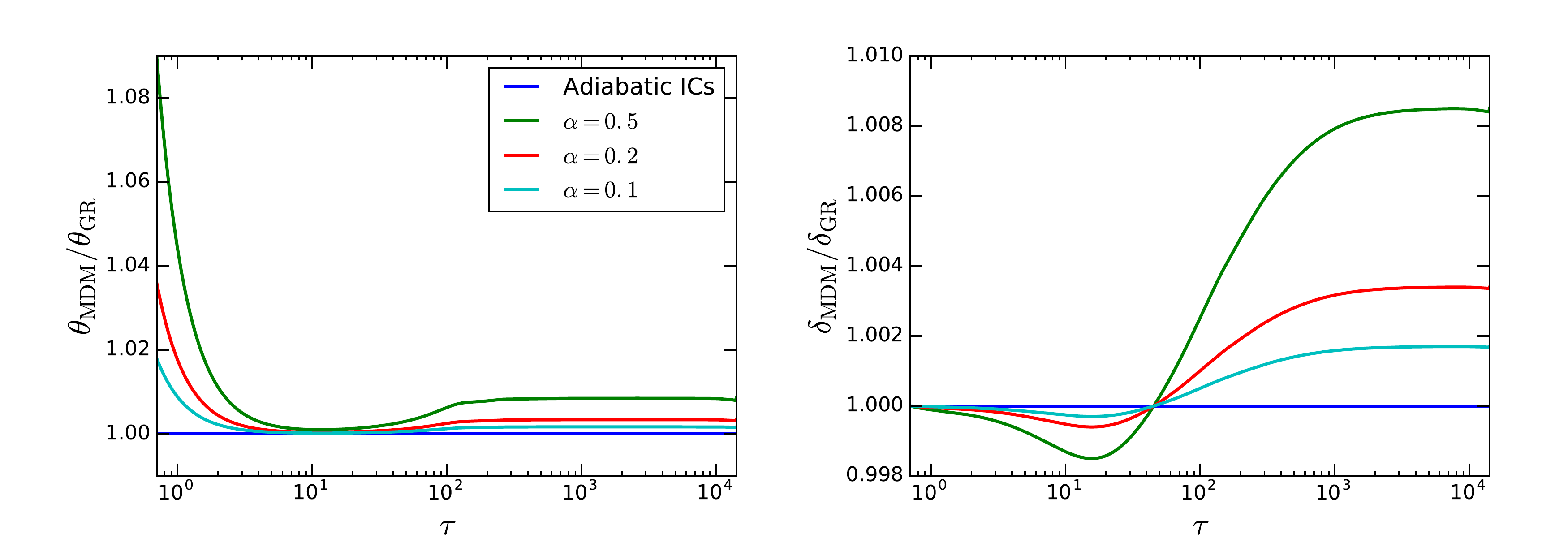}}
\caption{Ratio of the mimetic field velocity divergence perturbation (left panel) and mimetic field density perturbation (right panel) in MDM (with $V=0$)with respect to that of DM assuming $\Lambda$CDM for $k=0.1$ $h$Mpc$^{-1}$, as a function of conformal time.  We consider different values of the parameter $\alpha$ (color coded), which represents the maximum deviation from the adiabatic ICs case.}
\label{fig:evolution_of_perturbations}
\end{figure}

We compare the evolution of perturbations in $\Lambda$CDM and MDM in figure~\ref{fig:evolution_of_perturbations}. As we can notice, even when we perfectly match the overdensity perturbation in the MDM model to the one of DM in $\Lambda$CDM and we assume for them the same initial conditions at early times, we observe deviations at late times generated by different ICs in the velocity sector. Hence any small change in ICs only in the velocities will generate in turn larger changes in cosmological observables.

\begin{figure}[t]
\centerline{
\includegraphics[width=1.0\columnwidth]{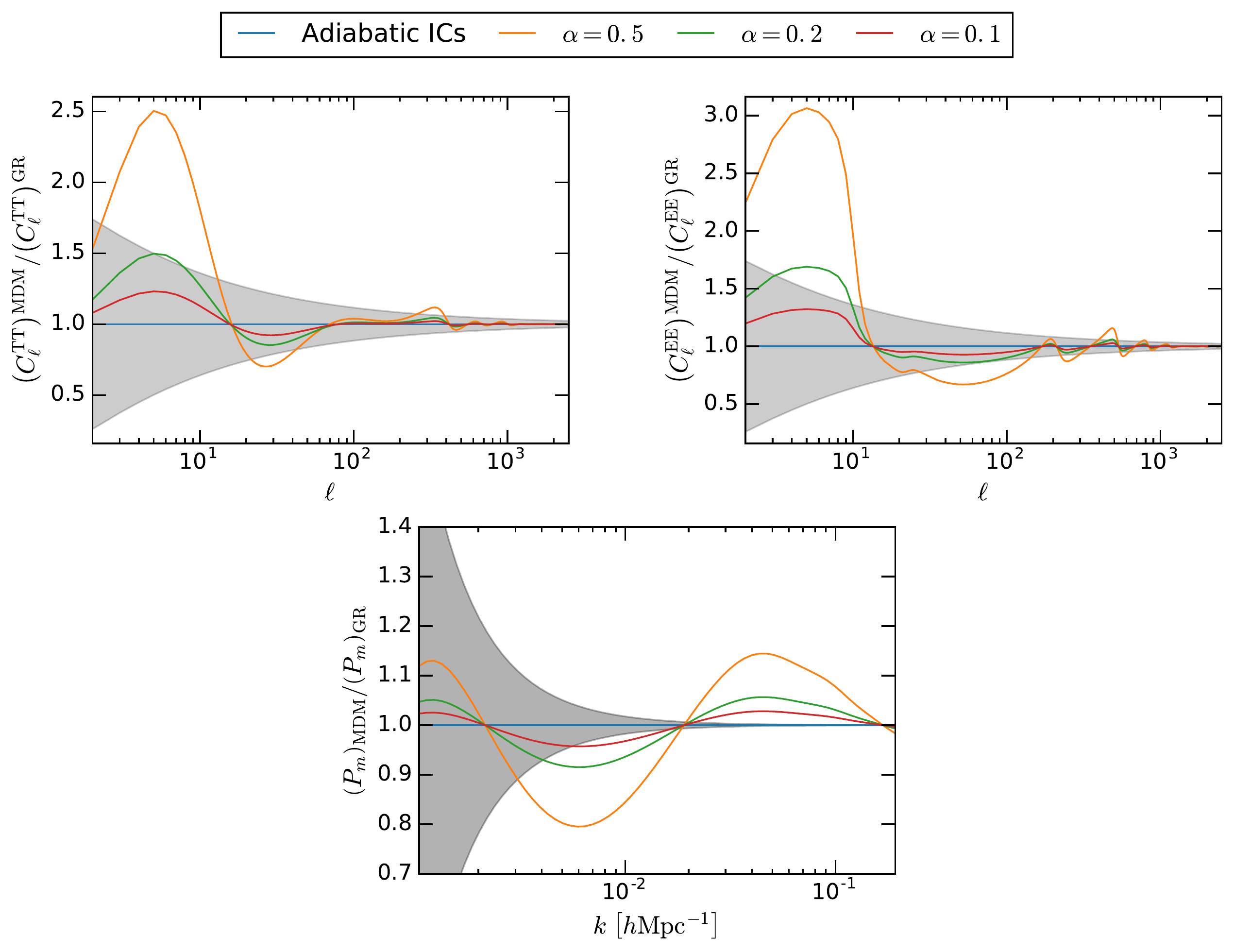}}
\caption{Ratio of the CMB temperature  angular power spectrum (upper left panel), the CMB E-mode polarization angular power spectrum (upper right panel) and the matter power spectrum at redshift $\bar{z}=1$ (bottom panel), in the MDM model (with $V=0$)with respect to the $\Lambda$CDM prediction, for the same cases considered in figure~\ref{fig:evolution_of_perturbations}. In the three panels the grey regions represents the cosmic variance limit (normalized  for $P_m$ as explained under equation~\eqref{eq:Pm_CV}).}
\label{fig:observables_in_mdm}
\end{figure}

These differences in the evolution of perturbations generate deviations in cosmological observables. We report them in figure~\ref{fig:observables_in_mdm} for the CMB temperature and polarization power spectra $(C^{TT}_\ell,\ C^{EE}_\ell)$ and for the matter power spectrum $(P_m)$. In all the cases, we analyse deviations up to $10\%$, $20\%$ and $50\%$ from adiabatic ICs, corresponding to $\alpha=0.1,\ 0.2,\ 0.5$, respectively.

We compare these deviations to cosmic variance uncertainty. For the angular power spectra, the cosmic variance reads as~\cite{Kamionkowski:1997na}
\begin{equation}
\frac{\sigma_{C_\ell}}{C_\ell} = \sqrt{\frac{2}{f_\mathrm{sky}(2\ell+1)}}
\end{equation} 
where $\ell$ is the multipole and $f_\mathrm{sky}$ is the observed fraction of the sky, independently from the chosen experiment. On the other hand, in the case of the matter power spectrum we have a dependence on the chosen survey, in fact the error is given by, see, e.g.,~\cite{audren:cosmicvariance},
\begin{equation}
\frac{\sigma_{P_m}}{P_m} = \sqrt{\frac{4\pi^2}{V_Sk^3\Delta\log k}}\left(1+\frac{1}{n_gP_m}\right),
\label{eq:Pm_CV}
\end{equation}
where $\Delta\log k$ is the bin size in $k$-space, $V_S$ is the volume of the survey and $n_g$ is the number density of galaxies. For an Euclid-like survey, with average redshift~$\bar{z}=1$, redshift bin width~$\Delta z=0.1$ and uniform binning of $\log k$, we have an estimated volume of~$V_S(\bar{z}) = 1.719\ \mathrm{Gpc}^3$ and number of galaxies of~$n_g(\bar{z})=1.998\times 10^{-3}\ \mathrm{Mpc}^{-3}$. Following Ref.~\cite{audren:cosmicvariance}, we further normalize the error bars to make them independent of the number of redshift bins and the width of the $k$ bins.

As can be seen from figure~\ref{fig:observables_in_mdm}, even fractional changes in the ICs generate deviations in the observable spectra which are detectable by cosmic variance limited experiments. Note that given the absence of a mechanism that automatically guarantees adiabatic ICs, a fractional change of less than $50\%$ represents a very modest variation. In the CMB temperature and polarization correlation functions, current observations by the Planck mission (cosmic variance limited up to $\ell \sim 2000$), rule out deviations larger than $20\%$ ($\alpha\gtrsim 0.2$) from the adiabatic initial conditions of $\Lambda$CDM. Furthermore, constraints by an Euclid-like mission will constrain any change at the percent level, as can be seen from the lower panel of figure~\ref{fig:observables_in_mdm}. By allowing also the~$\kappa$ parameter, $g$ function and its ICs, to vary we would expect much more significant deviations.

In other words, the free parameter~$\kappa$ and the ICs of the DM sector need to be fine-tuned at the $10\%$ level with current observations and they will be constrained at better than percent level with Euclid-like observations. Note that the fine-tuning problem is more severe than it looks as it is a function and not a simple number that needs to be adjusted to reproduce exactly adiabaticity (thus the level of fine-tuning extends to infinite degrees of freedom). Our main finding is that modifications of gravity that do not naturally produce a mechanism to generate adiabatic initial conditions do suffer from serious fine-tuning issues in the form of fine-tuning of free functions.


\section{Conclusions}
\label{sec:conclusions}
Despite its great success in describing the Universe we live in, the $\Lambda$CDM model does not provide any insight into what actually is the nature of its two main constituents, DM and dark energy. In this work we have explored in detail the predictions of a modified gravity model where the phenomenology associated to DM is described by pure geometry rather than elementary particles or compact objects. In particular, as a proof of principle, we focused on the mimetic dark matter model. 

After providing an alternative formulation to perturbation theory in this model, we found that this modified gravity model is naturally able to reproduce DM phenomenology, however it also contains free parameters and functions whose ICs need to be tuned in order to match observational data. Since the model does not naturally produce adiabatic initial conditions for the mimetic field, it requires extra tuning to reproduce observations. The purpose of this work was to highlight that this model, in its most basic form (i.e without a potential in the case of MDM), requires further development. In particular, to produce adiabatic initial conditions and, at the same time, to describe late time evolution of the universe, specific functions must be introduced at the level of the action. In other words, to reproduce observations, we do not have to tune only parameters, like in $\Lambda$CDM, but also the specific shape of functions.

We have modified the public Boltzmann code \texttt{CLASS} to compute both the evolution of perturbations and standard cosmological observables, as the matter power spectrum and cosmic microwave background temperature and polarization power spectrum, of the MDM model. Our modified version of \texttt{CLASS} is available on GitHub\footnote{\url{https://github.com/ark93-cosmo/CLASS_Modified_MDM}}. Several studies showed that ghosts and gradient instabilities may develop in an Universe filled only with the mimetic field~\cite{Ijjas:2016pad,Firouzjahi:2017txv}. In our numerical computations, which include matter and radiation, we did not impose by hand any extra stability requirement, hence we note that in a more realistic set-up this does not represent a problem of the model in its simplest version. However, if adding extra specific higher derivative couplings was needed to prevent these instabilities to emerge, as shown in refs.~\cite{Zheng:2017qfs,Hirano:2017zox,Gorji:2017cai}, we would need a higher level of fine tuning. We proved that current and future cosmic variance dominated experiments are able to detect small deviations from perfect adiabatic ICs, even in the conservative case where only the ICs of the velocity perturbations were allowed to vary by a small fraction. If all the free parameters and functions of the theory were allowed to deviate from its $\Lambda$CDM analogue, deviations would be much larger and would have been detected, for instance by Planck. 

We conclude by noticing that any modification of gravity that does not generically predict adiabatic ICs will suffer from severe fine tuning problems, since the degree of fine tuning for arbitrary functions is actually infinite and the model does not contain any attractor solution. This can be a route to restrict modifications of gravity and guide model building when abandoning General Relativity.


\section*{Acknowledgments}
ARK would like to thank Cyril Pitrou, Samuel Brieden and David Valcin for help and discussions. We thank Alexander Ganz, Justin Khoury, Sunny Vagnozzi and Alexander Vikman for useful comments on the draft. Funding for this work was partially provided by the Spanish MINECO under projects AYA2014-58747-P AEI/FEDER, UE, and MDM-2014-0369 of ICCUB (Unidad de Excelencia Mar\'ia de Maeztu). ARK is supported by project AYA2014-58747-P AEI/FEDER. NB is supported by the Spanish MINECO under grant BES-2015-073372. JLB was supported by the Spanish MINECO under grant BES-2015-071307, co-funded by the ESF, during most of the development of this work. Part of the calculations were done using the \texttt{xAct} package of Mathematica.



\begin{thebibliography}{10}

\bibitem{will:grstatusreview}
C.~M. Will, ``The Confrontation between General Relativity and Experiment'',
  \href{http://dx.doi.org/10.12942/lrr-2014-4}{{\em Living Reviews in
  Relativity} {\bfseries 17} no.~1, (Jun, 2014) 4},
  \href{http://arxiv.org/abs/1403.7377}{{\ttfamily arXiv:1403.7377}}.

\bibitem{Abbott:2016blz}
The {\bfseries LIGO Scientific Collaboration and Virgo Collaboration}, B.~P.
  Abbott {\em et~al.}, ``{Observation of Gravitational Waves from a Binary
  Black Hole Merger}'',
  \href{http://dx.doi.org/10.1103/PhysRevLett.116.061102}{{\em Phys. Rev.
  Lett.} {\bfseries 116} no.~6, (2016) 061102},
  \href{http://arxiv.org/abs/1602.03837}{{\ttfamily arXiv:1602.03837}}.

\bibitem{abbott:multimessenger}
B.~P. Abbott {\em et~al.}, ``Multi-messenger Observations of a Binary Neutron
  Star Merger'', \href{http://dx.doi.org/10.3847/2041-8213/aa91c9}{{\em The
  Astrophysical Journal} {\bfseries 848} no.~2, (Oct, 2017) L12},
  \href{http://arxiv.org/abs/1710.05833}{{\ttfamily arXiv:1710.05833}}.

\bibitem{abbott:o12catalog}
B.~P. Abbott {\em et~al.}, ``GWTC-1: A Gravitational-Wave Transient Catalog of
  Compact Binary Mergers Observed by LIGO and Virgo during the First and Second
  Observing Runs'', \href{http://arxiv.org/abs/1811.12907}{{\ttfamily
  arXiv:1811.12907}}.

\bibitem{eht:smbhdetection}
The {\bfseries EHT Collaboration}, ``First M87 Event Horizon Telescope Results.
  IV. Imaging the Central Supermassive Black Hole'',
  \href{http://dx.doi.org/10.3847/2041-8213/ab0e85}{{\em ApJL} {\bfseries 875}
  (2019) 4}, \href{http://arxiv.org/abs/1906.11241}{{\ttfamily
  arXiv:1906.11241}}.

\bibitem{Aghanim:2018eyx}
The {\bfseries Planck Collaboration}, N.~Aghanim {\em et~al.}, ``{Planck 2018
  results. VI. Cosmological parameters}'',
  \href{http://arxiv.org/abs/1807.06209}{{\ttfamily arXiv:1807.06209}}.

\bibitem{Alam_bossdr12}
The {\bfseries SDSS-III BOSS}, S.~{Alam} and et~al., ``{The clustering of
  galaxies in the completed SDSS-III Baryon Oscillation Spectroscopic Survey:
  cosmological analysis of the DR12 galaxy sample}'',
  \href{http://dx.doi.org/10.1093/mnras/stx721}{{\em MNRAS} {\bfseries 470}
  (Sept., 2017) 2617--2652}, \href{http://arxiv.org/abs/1607.03155}{{\ttfamily
  arXiv:1607.03155}}.

\bibitem{Bertone:2016nfn}
G.~Bertone and D.~Hooper, ``{History of dark matter}'',
  \href{http://dx.doi.org/10.1103/RevModPhys.90.045002}{{\em Rev. Mod. Phys.}
  {\bfseries 90} no.~4, (2018) 045002},
  \href{http://arxiv.org/abs/1605.04909}{{\ttfamily arXiv:1605.04909}}.

\bibitem{bertone:dmstatusreview}
G.~Bertone and T.~M.~P. Tait, ``A new era in the search for dark matter'',
  \href{http://dx.doi.org/10.1038/s41586-018-0542-z}{{\em Nature} {\bfseries
  562} (Oct, 2018) 51--56}, \href{http://arxiv.org/abs/1810.01668}{{\ttfamily
  arXiv:1810.01668}}.

\bibitem{Ratra}
B.~{Ratra} and P.~J.~E. {Peebles}, ``{Cosmological consequences of a rolling
  homogeneous scalar field}'',
  \href{http://dx.doi.org/10.1103/PhysRevD.37.3406}{{\em PRD} {\bfseries 37}
  no.~12, (Jun, 1988) 3406--3427}.

\bibitem{chamseddine:mimeticdm}
A.~H. Chamseddine and V.~Mukhanov, ``Mimetic dark matter'',
  \href{http://dx.doi.org/10.1007/JHEP11(2013)135}{{\em Journal of High Energy
  Physics} {\bfseries 2013} no.~11, (Nov, 2013) 135},
  \href{http://arxiv.org/abs/1308.5410}{{\ttfamily arXiv:1308.5410}}.

\bibitem{Golovnev:lagrangeMultiplier}
A.~Golovnev, ``{On the recently proposed Mimetic Dark Matter}'',
  \href{http://dx.doi.org/10.1016/j.physletb.2013.11.026}{{\em Phys. Lett.}
  {\bfseries B728} (2014) 39--40},
  \href{http://arxiv.org/abs/1310.2790}{{\ttfamily arXiv:1310.2790}}.

\bibitem{Barvinsky:ghost}
A.~O. Barvinsky, ``{Dark matter as a ghost free conformal extension of Einstein
  theory}'', \href{http://dx.doi.org/10.1088/1475-7516/2014/01/014}{{\em JCAP}
  {\bfseries 1401} (2014) 014},
  \href{http://arxiv.org/abs/1311.3111}{{\ttfamily arXiv:1311.3111}}.

\bibitem{Ganz:2018mqi}
A.~Ganz, P.~Karmakar, S.~Matarrese, and D.~Sorokin, ``{Hamiltonian analysis of
  mimetic scalar gravity revisited}'',
  \href{http://dx.doi.org/10.1103/PhysRevD.99.064009}{{\em Phys. Rev.}
  {\bfseries D99} no.~6, (2019) 064009},
  \href{http://arxiv.org/abs/1812.02667}{{\ttfamily arXiv:1812.02667}}.

\bibitem{chamseddine:mimeticcosmo}
A.~H. Chamseddine, V.~Mukhanov, and A.~Vikman, ``Cosmology with Mimetic
  Matter'', \href{http://dx.doi.org/10.1088/1475-7516/2014/06/017}{{\em Journal
  of Cosmology and Astroparticle Physics} {\bfseries 2014} no.~06, (2014) 017},
  \href{http://arxiv.org/abs/1403.3961}{{\ttfamily arXiv:1403.3961}}.

\bibitem{arroja:mimeticdisformallagrange}
F.~Arroja, N.~Bartolo, P.~Karmakar, and S.~Matarrese, ``The two faces of
  mimetic Horndeski gravity: disformal transformations and Lagrange
  multiplier'', \href{http://dx.doi.org/10.1088/1475-7516/2015/09/051}{{\em
  Journal of Cosmology and Astroparticle Physics} {\bfseries 2015} no.~09,
  (2015) 051}, \href{http://arxiv.org/abs/1506.08575}{{\ttfamily
  arXiv:1506.08575}}.

\bibitem{arroja:mimeticcosmoperturbations}
F.~Arroja, N.~Bartolo, P.~Karmakar, and S.~Matarrese, ``Cosmological
  perturbations in mimetic Horndeski gravity'',
  \href{http://dx.doi.org/10.1088/1475-7516/2016/04/042}{{\em Journal of
  Cosmology and Astroparticle Physics} {\bfseries 2016} no.~04, (2016) 042},
  \href{http://arxiv.org/abs/1512.09374}{{\ttfamily arXiv:1512.09374}}.

\bibitem{arroja:mimeticlss}
F.~Arroja, T.~Okumura, N.~Bartolo, P.~Karmakar, and S.~Matarrese, ``Large-scale
  structure in mimetic Horndeski gravity'',
  \href{http://dx.doi.org/10.1088/1475-7516/2018/05/050}{{\em Journal of
  Cosmology and Astroparticle Physics} {\bfseries 2018} no.~05, (2018) 050},
  \href{http://arxiv.org/abs/1708.01850}{{\ttfamily arXiv:1708.01850}}.

\bibitem{Ganz:2018vzg}
A.~Ganz, N.~Bartolo, P.~Karmakar, and S.~Matarrese, ``{Gravity in mimetic
  scalar-tensor theories after GW170817}'',
  \href{http://dx.doi.org/10.1088/1475-7516/2019/01/056}{{\em JCAP} {\bfseries
  1901} no.~01, (2019) 056}, \href{http://arxiv.org/abs/1809.03496}{{\ttfamily
  arXiv:1809.03496}}.

\bibitem{Sebastiani:2016ras}
L.~Sebastiani, S.~Vagnozzi, and R.~Myrzakulov, ``{Mimetic gravity: a review of
  recent developments and applications to cosmology and astrophysics}'',
  \href{http://dx.doi.org/10.1155/2017/3156915}{{\em Adv. High Energy Phys.}
  {\bfseries 2017} (2017) 3156915},
  \href{http://arxiv.org/abs/1612.08661}{{\ttfamily arXiv:1612.08661}}.

\bibitem{mirzagholi:imperfectdarkmatter}
L.~Mirzagholi and A.~Vikman, ``Imperfect Dark Matter'',
  \href{http://dx.doi.org/10.1088/1475-7516/2015/06/028}{{\em Journal of
  Cosmology and Astroparticle Physics} {\bfseries 2015} no.~06, (Jun, 2015)
  028}, \href{http://arxiv.org/abs/1412.7136}{{\ttfamily arXiv:1412.7136}}.

\bibitem{ramazanov:imperfectdarkmatter}
S.~Ramazanov, ``Initial conditions for imperfect dark matter'',
  \href{http://dx.doi.org/10.1088/1475-7516/2015/12/007}{{\em Journal of
  Cosmology and Astroparticle Physics} {\bfseries 2015} no.~12, (Dec, 2015)
  007}, \href{http://arxiv.org/abs/1507.00291}{{\ttfamily arXiv:1507.00291}}.

\bibitem{lim:constraintequation}
E.~A. Lim, I.~Sawicki, and A.~Vikman, ``Dust of dark energy'',
  \href{http://dx.doi.org/10.1088/1475-7516/2010/05/012}{{\em Journal of
  Cosmology and Astroparticle Physics} {\bfseries 2010} no.~05, (May, 2010)
  012}, \href{http://arxiv.org/abs/1003.5751}{{\ttfamily arXiv:1003.5751}}.

\bibitem{ma:cosmoperturbations}
C.-P. Ma and E.~Bertschinger, ``Cosmological Perturbation Theory in the
  Synchronous and Conformal Newtonian Gauges'',
  \href{http://dx.doi.org/10.1086/176550}{{\em Astrophysical Journal}
  {\bfseries 455} (Dec, 1995) 7},
  \href{http://arxiv.org/abs/astro-ph/9506072}{{\ttfamily
  arXiv:astro-ph/9506072}}.

\bibitem{bardeen:potentials}
J.~M. Bardeen, ``Gauge-invariant cosmological perturbations'',
  \href{http://dx.doi.org/10.1103/PhysRevD.22.1882}{{\em Phys. Rev. D}
  {\bfseries 22} (Oct, 1980) 1882--1905}.

\bibitem{lima:quartessence}
J.~Lima, J.~Cunha, and J.~Alcaniz, ``Simplified quartessence cosmology'',
  \href{http://dx.doi.org/10.1016/j.astropartphys.2009.01.004}{{\em
  Astroparticle Physics} {\bfseries 31} no.~3, (2009) 233 -- 236},
  \href{http://arxiv.org/abs/astro-ph/0611007}{{\ttfamily
  arXiv:astro-ph/0611007}}.

\bibitem{peiris:wmapng}
H.~V. Peiris, E.~Komatsu, L.~Verde, D.~N. Spergel, C.~L. Bennett, M.~Halpern,
  G.~Hinshaw, N.~Jarosik, A.~Kogut, M.~Limon, S.~S. Meyer, L.~Page, G.~S.
  Tucker, E.~Wollack, and E.~L. Wright, ``First-Year Wilkinson Microwave
  Anisotropy Probe (WMAP) Observations: Implications For Inflation'',
  \href{http://dx.doi.org/10.1086/377228}{{\em The Astrophysical Journal
  Supplement Series} {\bfseries 148} (Sep, 2003) 213--231},
  \href{http://arxiv.org/abs/astro-ph/0302225}{{\ttfamily
  arXiv:astro-ph/0302225}}.

\bibitem{Padilla:2019fju}
L.~E. Padilla, J.~A. V\'azquez, T.~Matos, and G.~Germ\'an, ``{Scalar Field Dark
  Matter Spectator During Inflation: The Effect of Self-interaction}'',
  \href{http://dx.doi.org/10.1088/1475-7516/2019/05/056}{{\em JCAP} {\bfseries
  1905} no.~05, (2019) 056}, \href{http://arxiv.org/abs/1901.00947}{{\ttfamily
  arXiv:1901.00947}}.

\bibitem{akrami:planckng2018}
The {\bfseries Planck Collaboration}, Y.~Akrami {\em et~al.}, ``Planck 2018
  results. X. Constraints on inflation'',
  \href{http://arxiv.org/abs/1807.06211}{{\ttfamily arXiv:1807.06211}}.

\bibitem{blas:class}
D.~Blas, J.~Lesgourgues, and T.~Tram, ``The Cosmic Linear Anisotropy Solving
  System ({CLASS}). Part {II}: Approximation schemes'',
  \href{http://dx.doi.org/10.1088/1475-7516/2011/07/034}{{\em Journal of
  Cosmology and Astroparticle Physics} {\bfseries 2011} no.~07, (Jul, 2011)
  034}, \href{http://arxiv.org/abs/1104.2933}{{\ttfamily arXiv:1104.2933}}.

\bibitem{Kamionkowski:1997na}
M.~Kamionkowski and A.~Loeb, ``{Getting around cosmic variance}'',
  \href{http://dx.doi.org/10.1103/PhysRevD.56.4511}{{\em Phys. Rev.} {\bfseries
  D56} (1997) 4511--4513},
  \href{http://arxiv.org/abs/astro-ph/9703118}{{\ttfamily
  arXiv:astro-ph/9703118}}.

\bibitem{audren:cosmicvariance}
B.~Audren, J.~Lesgourgues, S.~Bird, M.~G. Haehnelt, and M.~Viel, ``Neutrino
  masses and cosmological parameters from a Euclid-like survey: Markov Chain
  Monte Carlo forecasts including theoretical errors'',
  \href{http://dx.doi.org/10.1088/1475-7516/2013/01/026}{{\em Journal of
  Cosmology and Astro-Particle Physics} {\bfseries 2013} no.~1, (Jan, 2013)
  026}, \href{http://arxiv.org/abs/1210.2194}{{\ttfamily arXiv:1210.2194}}.

\bibitem{Ijjas:2016pad}
A.~Ijjas, J.~Ripley, and P.~J. Steinhardt, ``{NEC violation in mimetic
  cosmology revisited}'',
  \href{http://dx.doi.org/10.1016/j.physletb.2016.06.052}{{\em Phys. Lett.}
  {\bfseries B760} (2016) 132--138},
\href{http://arxiv.org/abs/1604.08586}{{\ttfamily arXiv:1604.08586 [gr-qc]}}.

\bibitem{Firouzjahi:2017txv}
H.~Firouzjahi, M.~A. Gorji, and S.~A. Hosseini~Mansoori, ``{Instabilities in
  Mimetic Matter Perturbations}'',
  \href{http://dx.doi.org/10.1088/1475-7516/2017/07/031}{{\em JCAP} {\bfseries
  1707} (2017) 031},
\href{http://arxiv.org/abs/1703.02923}{{\ttfamily arXiv:1703.02923 [hep-th]}}.

\bibitem{Zheng:2017qfs}
Y.~Zheng, L.~Shen, Y.~Mou, and M.~Li, ``{On (in)stabilities of perturbations in
  mimetic models with higher derivatives}'',
  \href{http://dx.doi.org/10.1088/1475-7516/2017/08/040}{{\em JCAP} {\bfseries
  1708} no.~08, (2017) 040},
\href{http://arxiv.org/abs/1704.06834}{{\ttfamily arXiv:1704.06834 [gr-qc]}}.

\bibitem{Hirano:2017zox}
S.~Hirano, S.~Nishi, and T.~Kobayashi, ``{Healthy imperfect dark matter from
  effective theory of mimetic cosmological perturbations}'',
  \href{http://dx.doi.org/10.1088/1475-7516/2017/07/009}{{\em JCAP} {\bfseries
  1707} no.~07, (2017) 009},
\href{http://arxiv.org/abs/1704.06031}{{\ttfamily arXiv:1704.06031 [gr-qc]}}.

\bibitem{Gorji:2017cai}
M.~A. Gorji, S.~A. Hosseini~Mansoori, and H.~Firouzjahi, ``{Higher Derivative
  Mimetic Gravity}'',
  \href{http://dx.doi.org/10.1088/1475-7516/2018/01/020}{{\em JCAP} {\bfseries
  1801} no.~01, (2018) 020},
\href{http://arxiv.org/abs/1709.09988}{{\ttfamily arXiv:1709.09988
  [astro-ph.CO]}}.

\end{thebibliography}

\providecommand{\href}[2]{#2}\begingroup\raggedright\endgroup

\end{document}